\documentclass[prb,twocolumn,aps,showpacs]{revtex4}
\usepackage{graphicx}% complex graphics
\usepackage{bm}% bold math
\usepackage{amssymb} %math symbols
\begin{document}

\title{Quantum site percolation on triangular lattice and
the integer quantum Hall effect}

\author{V. V. Mkhitaryan and M. E. Raikh}

\affiliation{ Department of Physics, University of Utah, Salt Lake
City, UT 84112}

\begin{abstract}
Generic {\em classical} electron motion in a strong perpendicular
magnetic field and random potential reduces to the {\em bond}
percolation on a {\em square} lattice. Here we point out that for
certain smooth 2D potentials with $120^{\circ}$ rotational
symmetry this problem reduces to the {\em site} percolation on a
{\em triangular} lattice.  We use this observation to develop an
approximate analytical description of the integer quantum Hall
transition. For this purpose we devise a quantum  generalization
of the real-space renormalization group (RG) treatment of the site
percolation on the triangular lattice. In quantum case, the RG
transformation describes the evolution of the {\em distribution}
of the $3\times 3$ scattering matrices at the sites. We find the
fixed point of this distribution and use it to determine the
critical exponent, $\nu$, for which we find the value $\nu \approx
2.3\div2.76$. The RG step involves only a {\em single} Hikami box,
and thus can serve as a minimal RG description of the quantum Hall
transition.
\end{abstract}


\maketitle

\section{Introduction}

Network-model formulation of the Anderson localization
problem was first introduced in Ref. \onlinecite{Shapiro}.
The key observation made in Ref. \onlinecite{Shapiro}
is that a complex motion of
electron in disorder potential can be reduced to the
 motion along the links of the network
(in both directions)  with disorder incorporated
via random phases of scattering from the
nodes of the network. Since then, the network-model
approach became a powerful tool for numerical studies
of disordered systems. In these studies the
randomness is incorporated into the phases accumulated
along the network links.
A great advantage of the
network-model approach to localization is that
it can be conveniently applied to various universality
classes\cite{Altland} of disorder. In order 
to  capture the specifics of a given class,
one has to impose an appropriate symmetry
requirements on the random phases on the links,
which fixes the form of $S$-matrix at the nodes.
This is achieved by introduction of an internal space
associated with the link of the network model, which
possesses a desired symmetry. For example, \cite{SO} with two-component
links (one component for one spin projection) for each direction
of propagation  the requirements of unitarity and time-reversal
symmetry allowed to reveal a delocalization transition expected
in two-dimensional systems with spin-orbit scattering.
Comprehensive review of the results obtained with
the help of the network
model is given in  Ref. \onlinecite{Ohtsuki}.
In addition to establishing the existence of localization transitions
in different classes, numerical simulations
of the network models with the help of transfer-matrix
method yields quantitative characteristics of
the critical point.
These characteristics include critical exponent,
critical level statistics, and
critical conductance distribution. Corresponding references
can be found in the review Ref. \onlinecite{Ohtsuki}.
Note that, with the exception of Ref. \onlinecite{Cardy}
where a particular two-channel network model was considered for
arbitrary graph, the underlying network in all previous
studies was a square lattice.

 Especially convenient for modeling with a network
is the {\em chiral} motion of a $2D$ electron in a disorder
potential and  a strong perpendicular magnetic field.
This is because the corresponding network is {\em directed}.
Directed character of the network with chiral scattering at
the nodes allowed Chalker and Coddington \cite{CC}
to demonstrate conclusively that there is only
a single delocalized state per Landau level.
This finding is of a great importance, since  it is the
underlying reason for sharp conductivity peaks at the quantum
Hall transition.
%%%%%%%%%%%%%%%%%%%
\begin{figure}[b]
\centerline{\includegraphics[width=85mm,angle=0,clip]{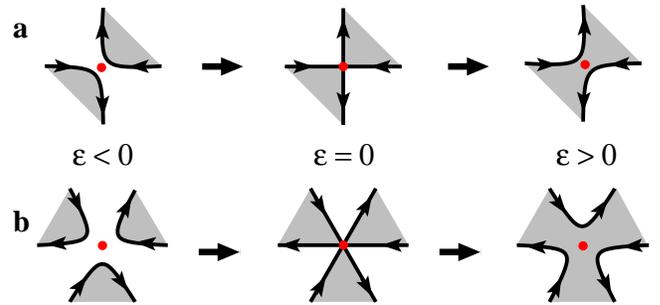}}
\caption{(Color online) (a) Evolution of equipotentials near the
saddle point of potential Eq. (\ref{Uxy}) as the energy
$\varepsilon$ crosses over from $\varepsilon<0$ to
$\varepsilon>0$; (b) evolution of equipotentials  near the nodes
of potential Eq. (\ref{BEHAVES}) with $\varepsilon$. Three
equipotential lines touch at $\varepsilon=0$.} \label{evolution}
\end{figure}
%%%%%%%%%%%%%%%%%%width=0.5\textwidth,

On  physical grounds,
the seminal  Chalker-Coddington
model \cite{CC} of the integer quantum Hall transition
can be introduced in a natural way with the help of a
two-dimensional potential
\begin{eqnarray}
\label{Uxy}
U(x,y)=\cos(\pi x)\cos(\pi y).
\end{eqnarray}
In this potential, the equipotential lines $U(x,y)=0$ form a
square lattice. For any nonzero $\varepsilon$ equipotentials
$U(x,y)=\varepsilon$ are closed. For positive $\varepsilon$ these
equipotentials encircle the maxima $(x,y)=(2m,2n)$ [and also
$(2m+1,2n+1)$] of $U(x,y)$, while for $\varepsilon<0$
equipotentials encircle the minima  $(x,y)=(2m,2n+1)$ [and
$(2m+1,2n)$]   of $U(x,y)$. In a strong perpendicular magnetic
field 2D electron drifts along equipotentials. Then
reconfiguration of equipotentials at $\varepsilon=0$, as
illustrated in Fig. \ref{evolution}a, manifests the change in the
character of motion.
%motion of a 2D electron in
%a strong perpendicular magnetic field,
%since this motion is a drift
%along the equipotentials.

Chalker and Coddington \cite{CC} captured the quantum character of
motion in $U(x,y)$ by assigning to the saddle points at
$(x_m,y_n)=(m-\frac{1}{2}, n-\frac{1}{2})$ where the potential
behaves as
\begin{eqnarray}
\label{behaves}
&&U(x-x_m,y-y_n)\approx\\
&& (-1)^{\scriptscriptstyle m+n}\pi^2(x-x_m)(y-y_n) =(-1)^{
\scriptscriptstyle m+n}\frac{\pi^2\rho^2}{2}\sin2\varphi,\nonumber
\end{eqnarray}
a scattering matrix
\begin{eqnarray}
\label{CCsc}
&&S_{cc}(\varepsilon)=\left(\begin{array}{cc}
\,\frac1{\sqrt{2}}+\varepsilon&\frac1{\sqrt{2}}-\varepsilon\\
\,\\
-\frac1{\sqrt{2}}+\varepsilon&\frac1{\sqrt{2}}+\varepsilon\\
\end{array}\right).
%\nonumber
\end{eqnarray}
Here $\rho$ and $\varphi$ are the polar coordinates with origin at
$(x_m,y_n)$, and $\varepsilon$ is the dimensionless energy.
For a realistic potential with magnitude, $U_0$, and correlation
length, $d$, dimensionless energy,  $\varepsilon$,
is the physical energy  measured in the units of
the width, $\Gamma \sim U_0l^2/d^2$, where, $l$,
is the magnetic length.
%%%%%%%%%%%%%%%%%%%
\begin{figure}[t]
\centerline{\includegraphics[width=80mm,angle=0,clip]{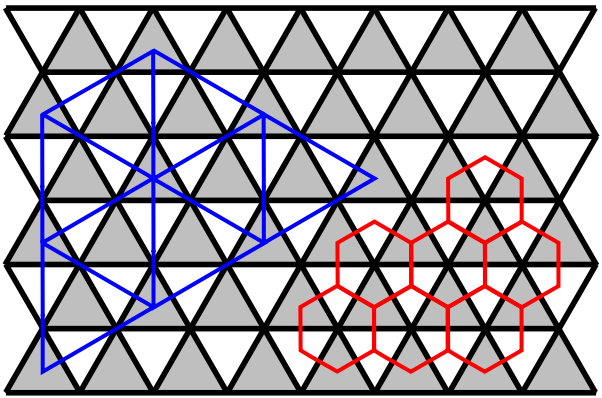}}
\caption{(Color online) Equipotential lines $\varepsilon=0$ of the
potential Eq. (\ref{BEHAVES}) constitute a triangular lattice.
Minima of $W(x,y)$ are in the centers of shaded triangles. Blue
lines constitute a triangular lattice of supersites with lattice
constant, $\sqrt{3}$. Red lines: hexagonal lattice with sites in
the centers of triangles. Each site of a triangular lattice is
surrounded by a hexagon. } \label{trilat}
\end{figure}
%%%%%%%%%%%%%%%%%%
Chalker and Coddington demonstrated that, in order to account for
a smooth disorder, it is sufficient to assume that the phases,
acquired by a drifting electron between the saddle points, are
random.

In this paper we show that the description of the
quantum Hall transition can also be obtained based
on the potential
\begin{eqnarray}
\label{BEHAVES}
W(x,y)=\sum_{n,m}V({\bf r}-n{\bf e}_1-m{\bf e}_2),
\end{eqnarray}
which has a {\em triangular} symmetry, i.e., ${\bf e}_1=(1,0)$ and
${\bf e}_2=(1/2,\sqrt{3}/2)$ are the basis unit vectors of a
triangular lattice; the function $V$ is defined inside a
``black half'' of the rhombus-shape unit cell, see  Fig. \ref{trilat},
in the following way
\begin{eqnarray}\label{tripot}
&&V(\rho,\varphi)=\left[\rho\cos\left(\varphi-\frac{\pi}6\right)-\frac1{2\sqrt{3}}\right]\\
&&\times\!\left[\rho\cos\left(\varphi-\frac{5\pi}6\right)\!-\frac1{2\sqrt{3}}\right]\!\!
\left[\rho\cos\left(\varphi-\frac{3\pi}2\right)\!-\frac1{2\sqrt{3}}\right]\!.\nonumber
\end{eqnarray}
Here $\rho$ and $\varphi$ are the polar coordinates
with respect to the origin at
$({\bf e}_1+{\bf e}_2)/3$. The form of potential in
the ``white half''
of a unit cell is given by Eq. (\ref{BEHAVES}) upon
replacement $V(\rho,\varphi)\rightarrow -V(\rho,\varphi-\pi/3)$
and shifting the origin to the center of the white triangle.

\begin{figure}[t]
\centerline{\includegraphics[width=90mm,angle=0,clip]{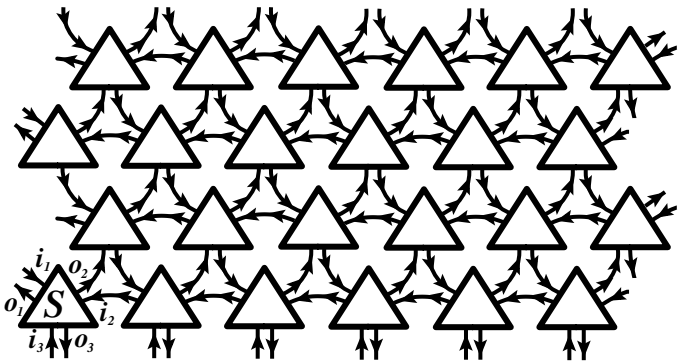}}
\caption{Network model on a triangular lattice is illustrated.
Three links enter a node and three links exit each node; the nodes
with $S$-matrices Eq. (\ref{sc}) are depicted as triangles.}
\label{network}
\end{figure}

Equipotentials of Eq.~(\ref{tripot}) evolve, as $\varepsilon$
passes through $0$, in a fashion {\it qualitatively different}
from the case of quadratic symmetry. As shown in Fig.
\ref{evolution}a, in the case of quadratic symmetry, black regions
(minima) get connected, while adjacent white regions (maxima) get
disconnected. By contrast, the behavior of the potential $W(x,y)$
near the nodes at ${\bf r}_{m,n}=(m-1/3){\bf e}_1+ (n-1/3){\bf
e}_2$ is given by
\begin{eqnarray}\label{bop}
W(\rho,\varphi)=-\rho^2
\sum_{q=0}^{\infty}c_q\sin\left[3(2q+1)\varphi\right],
\end{eqnarray}
where
%$c_m=\frac2{(2m+1)(9(2m+1)^2-4)}$.
$c_q=\frac{2\sqrt{3}}{\pi}\left[(2q+1)(9(2q+1)^2-4)\right]^{-1}$.
%$-\rho^2\sin(3\varphi)$.
Corresponding  evolution of
equipotentials is illustrated in Fig.~\ref{evolution}b. We see
that, as $\varepsilon$ crosses zero, {\em three} black regions get
joined at $\varepsilon=0$ {\em simultaneously}. This suggests that
quantum mechanical description of motion in the potential $W(x,y)$
requires, in addition to random phases on the links, introducing a
$3\times3$ scattering matrix at each node. Below we argue that the
form of this matrix is
\begin{eqnarray}
\label{sc}
&&S_{\vartriangle}(\varepsilon)=\left(\begin{array}{ccc}
\,\frac23(1+\varepsilon)&-\frac13&\,\frac23(1-\varepsilon)\\
%\,2/3&-1/3&\,2/3\\
\,\\
\,\frac23(1-\varepsilon)&\,\frac23(1+\varepsilon)&-\frac13\\
%\,2/3&\,2/3&-1/3\\
\,\\
-\frac13&\,\frac23(1-\varepsilon)&\,\frac23(1+\varepsilon)
%-1/3&\,2/3&\,2/3
\end{array}\right).
%\nonumber
\end{eqnarray}
With matrices Eq. (\ref{sc}) in the nodes, the corresponding
network model is shown in Fig. \ref{network}.
In
%the network Fig.~\ref{network}
this network the phases on the links are
random, as in Chalker-Coddington model, while all $S$-matrices in
the nodes, Eq. (\ref{sc}), are the same. As we demonstrate in
present paper, this model can be treated numerically using
the same MacKinnon-Kramer finite-size scaling algorithm
\cite{Kramer} that was employed in Ref. \onlinecite{CC} (for
subsequent numerical studies of the Chalker-Coddington model see
Refs.
\onlinecite{Kivelson93,LeeChalker94,LeeChalkerKo94,wen94,ruzin95,kagalovsky95,
kagalovsky97,wang96,fisher97,klesse,zirnbauer99,pryadko,klesse01}
and review articles \cite{Ohtsuki, reviews}).

Chalker-Coddington model can be viewed as
quantum version of the classical {\em bond}
percolation.
Establishing one to one correspondence between
the classical bonds and the links
is possible due to {\em directed} character of the
chiral CC network. On the other hand,
it was demonstrated in Ref. \onlinecite{Klein}
that a simple real-space renormalization group (RG)
procedure, based on decimation, leads to the closed equation
for the classical bond percolation threshold. This procedure
reproduces
the  exact threshold value and yields a very accurate estimate
for the critical exponent.
In Refs. \onlinecite{aram,arovas,cain,cain03,zulicke}
the classical RG procedure\cite{Klein} was generalized to
the quantum bond percolation. It was shown that corresponding
integral RG equation yields, in addition to the accurate
value of the quantum critical exponent, a very accurate
distribution of the critical conductance.
 A real-space
RG procedure  for a {\em site}  percolation
was proposed in the same paper Ref. \onlinecite{Klein}.
This procedure is simpler than for the bond percolation
and yields more accurate results. In the
present paper this procedure is extended to the quantum case,
where it describes the critical behavior of the
directed network Fig. \ref{network}.
As in the classical case, the quantum RG analysis of
on the triangular lattice is much easier than
the quantum RG analysis of the  Chalker-Coddington
on the square lattice.
This analysis is presented in Sections II-V.
%%%%%%%%%%%%%%%%%%%
%%%%%%%%%%%%%%%%%%

\section{RG procedure for classical percolation on a
triangular network}

%However,
In order to
develop the RG description, it is necessary to incorporate
disorder into the $S$-matrices. The reason is that at each RG-step
{\em three} $S$-matrices, connected by the links, are combined
into one {\em super}-$S$-matrix. Then the randomness in phases
translates into the randomness of the elements of {\em
super}-$S$-matrix.

A natural way to devise the quantum RG description is to start
from a classical problem of electron drift along the sides
of triangles,
%equipotential lines in
Fig. \ref{trilat}.

\subsection{Classical drift on triangular lattice}

Assume that the potential $W(x,y)$ is perturbed near the nodes as
\begin{equation}
\label{Random}
W(\rho,\phi)=V_0-c_0\rho^2\sin(3\varphi)-c_1\rho^2\sin(9\varphi)-\cdots,
\end{equation}
where the random shift, $V_0$, is much smaller than $1$, but much
bigger than the ``quantum'' energy width, $\Gamma$. Depending on
the shift, an electron drifting along equipotential
$W=\varepsilon$ towards a node, turns either to the left
($V_0<\varepsilon$)  or to the right ($V_0>\varepsilon$), see Fig.
\ref{evolution}b. We can conventionally call the  nodes with
$V_0<\varepsilon$ and $V_0>\varepsilon$ as ``black'' and ``white''
lattice sites, respectively. It is obvious that when the average,
$\langle V_0 \rangle$, is zero, the percolation threshold in the
potential Eq. (\ref{Random}) is $\varepsilon=0$. In the language
of sites, the same is to say that $50\%$ of sites are black at
$\varepsilon=0$. It is, in fact, well-known \cite{original}
that the exact
threshold for site percolation on a triangular lattice is $50\%$.
However, it is much less obvious that there is a {\em complete
equivalence} between electron drift in potential Eq.
(\ref{Random}) and the site percolation on a triangular lattice.
Superficially, this can be seen from the fact that when two
neighboring sites are black, they are connected via the black
region. Still, a rigorous proof requires additional steps, in
particular, introducing auxiliary hexagons, see Fig. \ref{trilat}.
This proof is given in the Appendix A.

\subsection{Classical RG scheme}

The above mapping on the percolation problem allows one  to employ
the RG approach to the site percolation on triangular lattice  put
forward in Ref. \onlinecite{Klein}. This procedure is much simpler
than the RG for the bond percolation  on the square lattice,
proposed in the same paper. Note in passing, that the bond
percolation  on the square lattice is the classical limit of the
Chalker-Coddington model \cite{aram}.

As shown in Fig. \ref{trilat}, at each RG step \cite{Klein} the
lattice constant increases by a factor of $\sqrt{3}$. A site of a
rescaled lattice,  a {\em supersite}, is either black or white
depending on the colors of the three constituting sites: if either
{\em all three} or {\em only two} out of three constituting sites
are black, then the supersite is black. Otherwise, the supersite
is white. Quantitatively, the probability $p^{\prime}$ for a
supersite to be black is expressed via the corresponding
probability for the original site as
\begin{equation}\label{cRG}
p^\prime=R(p)=p^3+3p^2(1-p).
\end{equation}
Fixed point, $p^{\prime}=p=p_c$, of the transformation
Eq.~(\ref{cRG}) reproduces the exact result $p_c=1/2$. Critical
exponent is determined by the condition that the correlation
radius, $\xi=(p-p_c)^{-\nu}$, on the original lattice is equal to
the correlation radius $\sqrt{3}\,(p^{\prime}-p_c)^{-\nu}$ on the
renormalized lattice, i.e.
\begin{equation}
\label{cNu}
\nu=\frac{\ln\left(\sqrt{3}\right)}{\ln\left(\frac{dp^\prime}{dp}
\right)_{p=p_c}}=\frac{\ln\left(\sqrt{3}\right)}
{\ln\left(3/2\right)}.
\end{equation}
Eq. (\ref{cNu}) yields  $\nu=1.354$, which differs from the exact
value, $\nu=4/3$, by only $1.6\%$.

The rationale behind the transformation Eq. (\ref{cRG}) is that
the supersite is located in the center of the black cell in Fig.
\ref{trilat}.  Then the color of the supersite reflects the
``percolation ability'' of this black triangular cell, so that,
even if one of the nodes, constituting the vertices of the
triangular cell is white, the cell still percolates ``over
black''.

\subsection{RG in the language of potential shifts}

At this point we make an observation that the above RG procedure
can be reformulated in the language of potential $W(x,y)$ with
random shifts in the nodes, $V_0$. Namely, for $V_0^{1}$,
$V_0^{2}$, and $V_0^{3}$ being the shifts at the nodes
constituting a supernode, the shift $V_0^{\prime}$ of the
supernode is defined as
\begin{equation}
\label{Mid} V_0^{\prime}=\mbox{Mid}\{V_0^{1},V_0^{2},V_0^{3}\},
\end{equation}
where Mid stands for  $V_0$ which is {\it smaller than maximal} but
{\it larger the minimal} out of the three numbers. With  $V_0^{\prime}$
defined by Eq. (\ref{Mid}), the RG equation Eq. (\ref{cRG})
describes the evolution of probability that the shift exceeds
$\varepsilon$.

The importance of the above observation is that reformulation of
classical RG procedure in terms of potential shifts opens a
possibility to capture the quantum-mechanical motion in the random
potential. A prescription  how to extend classical description to
the quantum case is \cite{aram}: the algorithm Eq. (\ref{Mid})
should be cast in the form of a {\em scattering problem}.

\section{Reformulation in terms of classical transmission}

We identify the scattering object as a point where three
equipotentials come close as  shown in Fig.~\ref{scatobj}.
Incident electron, $i_1$, either proceeds along the same
equipotential into $o_1$ (reflection) or switches equipotentials
and proceeds along $o_2$. Retention of equipotential (reflection)
corresponds to positive $V_0-\varepsilon$ in the vertex, encircled
in Fig.~\ref{scatobj}. In terms of scattering matrix
\begin{eqnarray}
\label{ss1}
&&\left(\begin{array}{c} o_1\\
o_2\\ o_3
\end{array}\right) = S\left(\begin{array}{c} i_1\\
i_2\\ i_3
\end{array}\right),
\end{eqnarray}
the same simple notion can be reformulated as follows. For positive
$V_0-\varepsilon$, the matrix $S=S_{+}$ is a {\em unit matrix}, while
for negative $V_0-\varepsilon$ the matrix $S=S_{-}$ has the form
\begin{eqnarray}
\label{s-}
&&S_-=\left(\begin{array}{ccc}
0&0&1\\
1&0&0\\
0&1&0
\end{array}\right).\nonumber
\end{eqnarray}
Superscattering object consists of three scattering objects, and
is shown in the same figure.

Now we reformulate Eqs. (\ref{cRG}) and (\ref{Mid}) in  yet another
language of $S$-matrices. Namely, the $S$-matrix of a superscattering
object is expressed via $S$-matrices  of constituting scattering
objects upon reducing the number of legs from $12$ to $6$.
We emphasize that this reduction can be carried out in two distinct
ways, as illustrated in Fig.~\ref{contraction}. The first way is
to perform contractions as $o_2\leftrightarrow i_2$,
$o_4\leftrightarrow i_4$, and $o_6\leftrightarrow i_6$. The second
variant of contractions is $o_2\leftrightarrow i_1$,
$o_4\leftrightarrow i_3$, and $o_6\leftrightarrow i_5$. Now it is
straightforward to check that RG transformation Eq.~(\ref{Mid})
corresponds to the following rule for $S$-matrix of the
superscattering object, $\tilde{S}$. If the $S$-matrices of either
all three or only two of constituting objects are $S_{+}$, then
$\tilde{S}=S_{+}$. In all other realizations, when at least two of
constituting objects have the matrix $S_{-}$, we have
$\tilde{S}=S_{-}$.

It is important to note that the above rule for $\tilde{S}$
applies {\em independently} of the way in which the contractions
in Fig.~\ref{contraction} are performed. This is not the case in
the quantum version to which we now turn.

\section{quantum generalization}
\begin{figure}[t]
\centerline{\includegraphics[width=80mm,angle=0,clip]{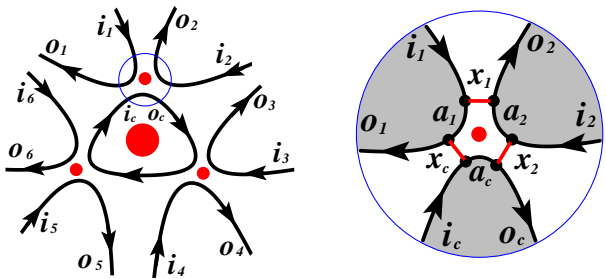}}
\caption{(Color online) Left: superscattering object is shown with
a big red dot, while constituting scattering objects are shown
with small red dots.  Right: detailed structure of the scattering
object; black dots show the points of the close contact of
equipotentials. Electron incident along equipotential, $i$, can
either proceed along the same equipotential, $a$, or swich
equipotential with probability $|x|^2$.  The latter processes are
illustrated with red dashes.} \label{scatobj}
\end{figure}

Quantum $S$-matrix of the scattering object differs from $S_{+}$
and $S_{-}$ in two respects. Firstly, at the points of close
contact between each pair of equipotentials electron can switch
equipotential {\em even for positive} $V_0-\varepsilon$, when it
is forbidden classically. Corresponding classically forbidden
transitions are illustrated in Fig. \ref{scatobj} with red dashes.
Secondly, upon travelling between two subsequent points of close
contact, electron accumulates the Aharonov-Bohm phase, $\phi_i$.
For example, the phase $\phi_1$ is accumulated in course of drift
between $i_1$ and $o_1$.  These phases are irrelevant in the
classical limit when the reflection amplitudes, $y_i$, are either
$0$ or $1$. However, for intermediate $0<y_i<1$ the amplitude for
an electron to execute a {\it close contour} around the center in
Fig. \ref{scatobj} (following the red dashes in the clockwise
direction) is {\em finite}. As a result, $\phi_i$ enter into
quantum scattering matrix. Explicit form of $S$ in terms of
$0<y_i<1$ and $\phi_i$ can be obtained by solving three pairs of
linear equations, describing quantum scattering at each of three
points of the close contact of equipotentials. We have
\begin{eqnarray}
\label{sm}
&&\hspace{-0.4cm}S=\frac{-1}{1-x_1x_2x_3e^{\scriptscriptstyle i\psi}}\times\\
&&\hspace{-0.4cm}\left(\begin{array}{ccc}
y_1y_3e^{\scriptscriptstyle i\phi_1} &x_1y_2
y_3e^{\scriptscriptstyle i(\phi_1+\phi_2)}
&x_1x_2e^{\scriptscriptstyle i\psi}-x_3\\
\,\\
x_2x_3e^{\scriptscriptstyle i\psi}-x_1
&y_1y_2e^{\scriptscriptstyle i\phi_2}
&y_1x_2y_3e^{\scriptscriptstyle i(\phi_2+\phi_3)}\\
\,\\
y_1y_2x_3e^{\scriptscriptstyle i(\phi_1+\phi_3)}
&x_1x_3e^{\scriptscriptstyle i\psi}-x_2
&y_2y_3e^{\scriptscriptstyle i\phi_3}
\end{array}\right),\nonumber
\end{eqnarray}
where
%$x_j$,
$x_j=\sqrt{1-y_j^2}$ stand for the transmission amplitudes, and
\begin{eqnarray}
\label{net}
\psi=\phi_1+\phi_2+\phi_3
\end{eqnarray}
is the net phase accumulated along the closed contour. It is easy
to check that the matrix Eq. (\ref{sm}) is unitary. It is also
straightforward to verify that in the classical limits, when {\em
all} $x_j=0$ or $x_j=1$, Eq. (\ref{sm}) correctly reproduces $S_+$
and  $S_-$, respectively. Detailed derivation of the form
Eq.~(\ref{sm}) of the scattering matrix is presented in Appendix
B.
%%%%%%%%%%%%%%%%%%%
%%%%%%%%%%%%%%%%%%

\subsection{The form of the matrix $S_{\vartriangle}$}

From  Eq.~(\ref{sm}) we can establish the form of the scattering
matrix $S_{\vartriangle}(\varepsilon)$, Eq.~(\ref{sc}), with the
help of the following duality argument. Due to triangular symmetry
of the potential $\rho^2\sin(3\varphi)$ we have $x_1=x_2=x_3=x$.
Consider now the transition point, $\varepsilon=0$. At this point
the probability for  electron incident along, say, $i_1$ (see Fig.
\ref{scatobj}) to be deflected to the left (along $o_1$) is equal
to the probability to be deflected to the right (along $o_2$).
Less trivial is that the phase, $\psi$, must be zero at
$\varepsilon=0$. This is the consequence of the fact that the
scattering problems for electron with energy $\varepsilon$ and
$-\varepsilon$ are equivalent if we change the drift direction
from clockwise to anti-clockwise; this change implies also the
change in the sign of $\psi$.

With $\psi=0$, the condition, $(1-x^2)^2=(x^2-x)^2$, of ``equal
deflection'' to the left and to the right, yields a single
physical root $x=-1/2$. Substituting it back in the matrix Eq.
(\ref{sm}) reduces it to the scattering matrix Eq. (\ref{sc}) with
$\varepsilon=0$. Then for probabilities $i_1\rightarrow o_1$ and
$i_1\rightarrow o_2$ we get $4/9$, while the probability
$i_1\rightarrow o_3$ is $1/9$. Including small finite
$\varepsilon$ can be also performed with the help of the duality
argument, namely that the probability of deflection to the left
with energy $\varepsilon$ is equal to the probability of
deflection to the right with energy $-\varepsilon$. On the other
hand, the change of probability $i_1\rightarrow o_3$ with
$\varepsilon$ is $\propto \varepsilon^2$.
%%%%%%%%%%%%%%%%%%%
\begin{figure}[t]
\centerline{\includegraphics[width=80mm,angle=0,clip]{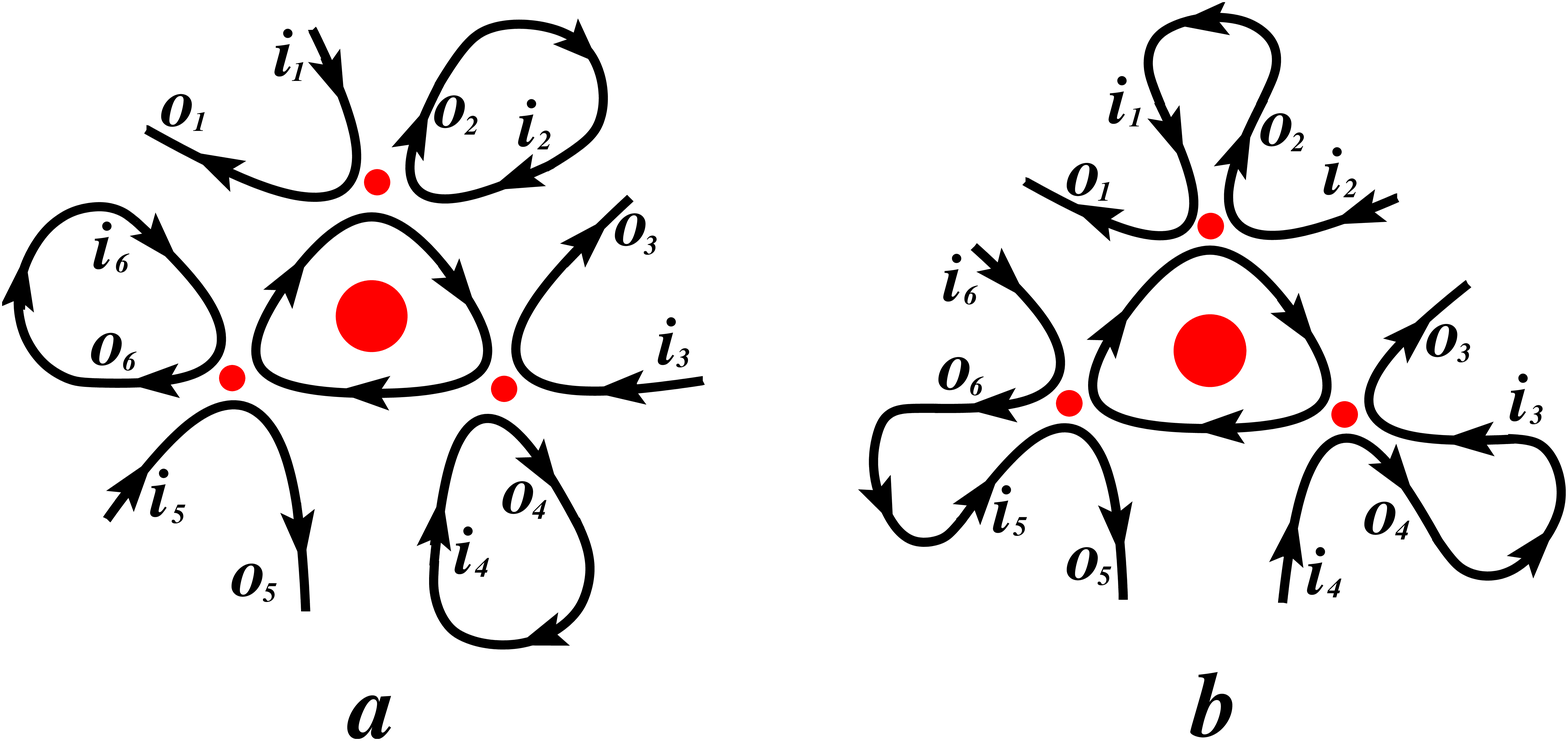}}
\caption{(Color online) Two contributions to the kernel Eq.
(\ref{kernel}) originating from two variants of contraction of
equipotentials are illustrated; (a) corresponds to $\tilde{x}^a$,
Eq. (\ref{i2o2}), while (b) corresponds to $\tilde{x}^b$,  Eq.
(\ref{o2i1}).} \label{contraction}
\end{figure}
%%%%%%%%%%%%%%%%%%

The role of the matrix $S_{\vartriangle}$ is central to the
transfer-matrix treatment of the network model; see Appendix C.

\subsection{Quantum RG equation}

We now turn to the quantum RG procedure. In contrast to evolution
of probability, $p$, upon rescaling of the lattice constant in the
classical case, this procedure is formulated in terms of evolution
of the {\em distribution function}, $Q(x)$, of the {\em absolute
values} of the transmission amplitudes, $x_i$. Thus the quantum
generalization of Eq. (\ref{cRG}) at the step $n$ is the following
recurrence relation
\begin{eqnarray}
\label{qRG}
&&Q_{n+1}(x)=\text{\large T}\left\{Q_n(x)\right\}\\
&&=\int\left(\prod\limits_{j=1}^3 dx_jQ_n(x_j)\right)
K(x,x_1,x_2,x_3).\nonumber
\end{eqnarray}
The kernel, $K$, represents the conditional probability that,
after performing the contractions, the transmission coefficient
$i_c\rightarrow o_1$ in Fig.  \ref{scatobj} is equal to $x$,
provided that the constituting transmission coefficients are
$x_1$, $x_2$, and $x_c=x_3$, as illustrated in Fig. \ref{scatobj}.
In analytical evaluation of the dependence $x(x_1,x_2,x_3)$ it is
important to take into account that this dependence is different
for two variants of contractions. For a variant
$o_2\leftrightarrow i_2$, shown in Fig. \ref{contraction}a, the
transmission coefficient is given by
\begin{eqnarray}
\label{i2o2}
&&\hspace{-0.3cm}|\tilde{x}_c^a|^2=\\
&&1-{\Bigg |}\frac{\left(\sqrt{1-x_1^2}\,
e^{i\varphi}+ \sqrt{1-x_2^2}\right)\sqrt{1-x_3^2}}{(1-x_1x_2x_3
e^{i\psi})e^{i\varphi}+\sqrt{(1-x_1^2)(1-x_2^2)}}{\Bigg
|}^2,\nonumber
\end{eqnarray}
where $\varphi$ is the phase along the contour ($o_2, i_2$), which
is now {\em closed}. Correspondingly, for the second kind of
closing $o_2\leftrightarrow i_1$, Fig. \ref{contraction}b, the
transmission coefficient has the form
\begin{eqnarray}
\label{o2i1}
&&\hspace{-0.3cm}|\tilde{x}_c^b|^2=\\
&&1-{\Bigg |}\frac{(x_1-
e^{i\varphi})\sqrt{(1-x_2^2)(1-x_3^2)}}{(1-x_1x_2x_3
e^{i\psi})e^{i\varphi} +x_2x_3e^{i\psi}-x_1}{\Bigg |}^2,\nonumber
\end{eqnarray}
where $\varphi$ stands for the phase along $(o_2, i_1)$. For
details of derivation of Eqs. (\ref{i2o2}), (\ref{o2i1}) see
Appendix B.
%%%%%%%%%%%%%%%%%%%
\begin{figure}[t]
\centerline{\includegraphics[width=90mm,angle=0,clip]{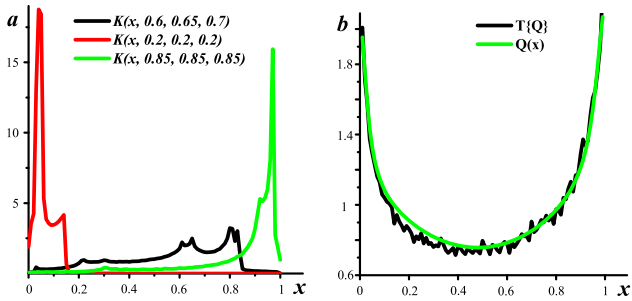}}
\caption{(Color online) Left: Kernel $K(x,x_1,x_2,x_3)$ is plotted
from Eq. (\ref{kernel}) for three sets $(x_1,x_2,x_3)$. Right:
approximate analytical solution (green line) of the RG equation
Eq.~(\ref{fp}), given by Eq.~(\ref{fpsol}), is plotted together
with r.h.s. of Eq.~(\ref{fp}), which is $\text{\large
T}\left\{Q(x)\right\}$ (black line).} \label{RGsol}
\end{figure}
%%%%%%%%%%%%%%%%%%

Relations Eqs. (\ref{i2o2}), (\ref{o2i1}) define also the
dependencies $\tilde{x}_1^{a,b}(x_1,x_2,x_3)$ and
$\tilde{x}_2^{a,b}(x_1,x_2,x_3)$. It is important that the central
phase, $\psi$, is {\it common} for all three dependencies
$\{\tilde{x}_j^a(\{x_i\},\varphi_j,\psi)\}_{\scriptscriptstyle
j=1,2,3}$ and $\{\tilde{x}_j^b(\{x_i\},\varphi_j,\psi)\}
_{\scriptscriptstyle j=1,2,3}$. A crucial step in extending
classical RG procedure to the quantum case is to follow the
"classical" prescription to choose a {\em middle} out of three
coefficients
\begin{eqnarray}
\label{alg} &&\hspace{-0.4cm}\tilde{x}^{a}(\{x_i\},\{\varphi_i\},\psi)=\\
&&\hspace{-0.4cm}\mbox{Mid}\Bigl (\tilde{x}_1^{a}(\{x_i\},\varphi_1,\psi),\,
\tilde{x}_2^{a}(\{x_i\},\varphi_2,\psi),\,
\tilde{x}_3^{a}(\{x_i\},\varphi_3,\psi)\Bigr),\nonumber
\end{eqnarray}
and the same for $\tilde{x}^{b}(\{x_i\},\{\varphi_i\},\psi)$. Eq.
(\ref{alg}) is a quantum generalization of the classical Eq.
(\ref{Mid}). Note that selection of middle value in Eq.
(\ref{alg}) is performed for {\em given} values of random phases,
$\varphi_1$, $\varphi_2$, $\varphi_3$, and $\psi$.  Within RG
procedure different phases are uncorrelated, and in evaluation of
the kernel we average over each of four phases independently.
Finally, taking into account that contractions $a$ and $b$ are
statistically equivalent, the expression for $K(x, \{x_i\})$
acquires the form
\begin{eqnarray}
\label{kernel} &&\hspace{-0.3cm}K(x, \{x_i\})=
\frac12{\Big\langle}\delta(x-\tilde{x}^a(\{x_i\},\{\varphi_i\},\psi))
{\Big\rangle}_{\scriptscriptstyle \{\varphi_i,\psi\}}\nonumber\\
&&+\frac12{\Big\langle}\delta(x-\tilde{x}^b(\{x_i\},\{\varphi_i\},\psi))
{\Big\rangle}_{\scriptscriptstyle \{\varphi_i,\psi\}} .
\end{eqnarray}
Quantum delocalization corresponds to the  fixed point of the
transformation Eq. (\ref{qRG}). It is found in the next Section.

\section{Numerical results for fixed point and critical exponent}

\subsection{The kernel}
The examples of the kernel, $K(x)$, calculated using Mathematica
from Eqs. (\ref{i2o2})-(\ref{kernel}), are plotted in
Fig.~\ref{RGsol} for different sets $x_1$, $x_2$, $x_3$. It is
seen that the kernel supports the attractive critical points
$x_1=x_2=x_3=0$ and $x_1=x_2=x_3=1$. Indeed, for values
$x_1=x_2=x_3=0.2$, the kernel is centered at even smaller value
$x\approx 0.04$, while for $x_1=x_2=x_3=0.85$ it is around bigger
value $x\approx 0.95$. In both cases the kernel is narrow. This is
because for ``classical'' transmission coefficients interference
does not play a role, so that the phases drop out from Eqs.
(\ref{i2o2}), (\ref{o2i1}). However, for intermediate values
$x_1=0.6$, $x_2=0.65$, $x_3=0.7$ the kernel extends over entire
interval $0<x<1$. It also exhibits peaks at $x= 0.03, 0.61, 0.65,
0.8$ and $0.83$. The origin of these peaks is the anomalous
contributions of phases $\varphi_0=0,\pi$ and $\psi_0=0,\pi$ to
the kernel. Indeed, for these values of phases we have $\partial
\tilde{x}^{a,b}/\partial\varphi =
\partial \tilde{x}^{a,b}/\partial\psi =0$, so that
\begin{eqnarray}
&&\tilde{x}^{\scriptscriptstyle a,b}(\varphi,\psi)=
\tilde{x}^{\scriptscriptstyle a,b}(\varphi_0,\psi_0)+
\frac{\partial^2_\varphi\tilde{x}^{\scriptscriptstyle
a,b}}2\!{\bigg |}_{\scriptscriptstyle
\varphi_0,\psi_0}\!\!\!\!(\varphi-\varphi_0)^2\\
&&+\frac{\partial^2_\psi\tilde{x}^{\scriptscriptstyle
a,b}}2\!{\bigg |}_{\scriptscriptstyle
\varphi_0,\psi_0}\!\!\!\!(\psi-\psi_0)^2
+\partial^2_{\varphi\psi}\tilde{x}^{\scriptscriptstyle
a,b}\!{\bigg |}_{\scriptscriptstyle
\varphi_0,\psi_0}\!\!\!\!(\varphi-\varphi_0)(\psi-\psi_0).\nonumber
\end{eqnarray}
It is easy to check that, when the quadratic form of second
derivatives is {\it negatively defined}, the corresponding
contribution to the kernel is $\propto \ln\vert
x-\tilde{x}^{a,b}(\varphi_0,\psi_0)\vert$. Therefore, the peaks in
the kernel reflect the fact that two loops corresponding to phases
$\phi$ and $\psi$ are insufficient for complete averaging. Still,
the phase volume of the singular contributions is small, so that
the fixed point of the transformation Eq. (\ref{qRG}) is not
sensitive to these singularities.
%%%%%%%%%%%%%%%%%%%
\begin{figure}[t]
\centerline{\includegraphics[width=90mm,angle=0,clip]{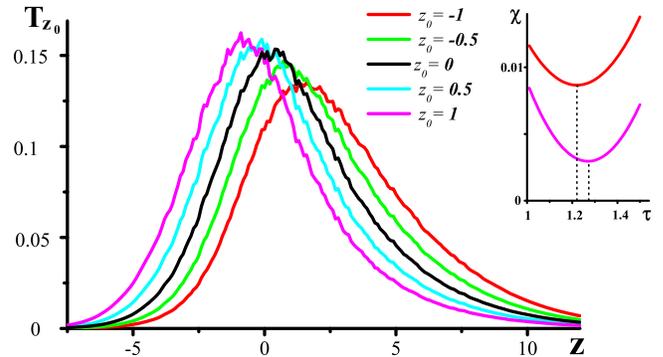}}
\caption{(Color online) $\text{\large T}_{z_0}=\text{ \large T}
\!\left\{\tilde{Q}(z-z_0)\right\}$ is plotted for different $z_0$.
Inset shows the variance, $\chi(\tau)$, plotted from Eq.
~(\ref{variance}). The minima correspond to $\tau=1.27$ and
$\tau=1.22$ for $z_0=1$ and $z_0=-1$, respectively.} \label{diffz}
\end{figure}
%%%%%%%%%%%%%%%%%%

\subsection{Fixed point}
The fixed point, $Q(x)$, of the quantum RG equation
Eq.~(\ref{qRG}), satisfies the following nonlinear integral equation
\begin{eqnarray}
\label{fp} Q(x)=\!\!\text{ \large T}\!\left\{Q(x)\right\}=
\!\!\int\!\!\left(\prod\limits_{j=1}^3 dx_jQ(x_j)\!\right)
\!\! K(x,x_1,x_2,x_3),\nonumber\\
\end{eqnarray}
which we have solved using Mathematica. Starting from initial
distribution, $Q(x)=1$, and performing analytical fit at each step
of successive approximations, the following expression for the
fixed point was obtained after the fourth step
\begin{eqnarray}
\label{fpsol} Q(x)=
%\frac{0.9}{1+1000x^2}+ \frac{0.9}{1+500(1-x)^2}\nonumber\\
\frac{0.9}{1+10^3x^2}+ \frac{0.9}{1+5\cdot 10^2(1-x)^2}\nonumber\\
+0.63x^3+0.55(1-x)^3+0.6.
\end{eqnarray}
This expression is plotted in Fig.~\ref{RGsol}b with a green line.
One can judge on the accuracy of the approximate solution
Eq.~(\ref{fpsol}) by substituting it into the right-hand side of
Eq.~(\ref{fp}). The result, black line in Fig.~\ref{RGsol}b, is
indeed very close to Eq.~(\ref{fpsol}).

The fixed point solution rises upon approaching $x=0$
and $x=1$. This behavior is inherited from
the classical percolation. Note that similar behavior
was found in Refs. \onlinecite{aram,cain} for the fixed point
of the quantum bond percolation on the square lattice.
Direct comparison with conductance distribution $P(G)$
found in  Refs. \onlinecite{aram,cain} can be performed
using the relation $P(G)=Q(\sqrt{G})/2\sqrt{G}$. This
comparison indicates that, while numerically the
fixed point distributions are close, Eq. (\ref{fpsol})
favors small values of $x$. Qualitatively, this
reflects the fact that at critical energy,
$\varepsilon=0$, electron incident along $i_1$
(see  Fig.  \ref{scatobj}) is more likely to
proceed along $a_1$ rather than switch to  $o_2$.
The asymmetry is seen more clearly if one interprets
$Q(x)$ in terms of distribution, $\tilde{Q}$ of heights of
the effective saddle point. This height is determined by
the relation $x^2=1/(e^z+1)$, so that
\begin{eqnarray}
\label{tildQ}
\tilde{Q}(z)=\frac{e^z}{2(e^z+1)^{3/2}}
Q\left(\left[e^z+1\right]^{-1/2}\right).
\end{eqnarray}
The distribution $\tilde{Q}(z)$ is shown in Fig. \ref{diffz} with
a black line. It has an asymmetry towards large $z$.

\subsection{Critical exponent}
To estimate the critical exponent, $\nu$, which governs the
divergence of localization radius, $\xi(\varepsilon) \propto
1/\varepsilon^{\nu}$, as a function of energy, $\varepsilon$, we
used the reasoning from  Refs.~\onlinecite{aram,cain}. Electron
with finite energy, $\varepsilon \ll \Gamma$, ``sees'' the shifted
distribution of the saddle point heights $\tilde{Q}(z-z_0)$, where
$z_0\ll 1$ is proportional to $\varepsilon$. The key step of the
reasoning \cite{aram,cain} is that, upon the RG transformation,
the electron travels on the lattice with the lattice constant
$\sqrt{3}$ and sees the shifted distribution of heights, $\text{
\large T}\!\left\{\tilde{Q}(z-z_0)\right\} =\tilde{Q}(z-\tau
z_0)$, where $\tau$ is some constant independent of $z_0$. After
subsequent $n$ RG steps this distribution evolves into
$\tilde{Q}(z-\tau^nz_0)$. When the shift accumulates to reach
$\sim 1$, the electron becomes localized within the size of a unit
cell of renormalized lattice. Then from the relations
\begin{eqnarray}
\label{reasoning}
\xi(z_0)\propto\frac1{z_0^\nu},\quad\xi=\bigl(\sqrt{3}\,\bigr)^{n},
\quad z_0\tau^n\sim1,
\end{eqnarray}
we find
\begin{eqnarray}
\label{nu}
\nu=\frac{\ln\sqrt{3}}{\ln\tau}.
\end{eqnarray}
This definition of $\nu$ is a quantum generalization of
Eq.~(\ref{cNu}). In Fig.~\ref{diffz} the result of calculation
$\text{ \large T}\!\left\{\tilde{Q}(z-z_0)\right\}$ for four
$z_0=-1$, $-0.5$, $0.5$, $1$ is shown. We see that for these $z_0$
the shape of $\text{ \large T}\!\left\{\tilde{Q}\right\}$ is only
slightly affected by the shift. The curves are approximately
equidistant, so that an estimate of $\tau$ can be obtained simply
from the horizontal separation $\approx 0.4$ between the
neighboring curves. This yields $\tau \approx 1.25$ and,
correspondingly, $\nu \approx 2.46$. For more accurate estimate we
studied the variance,
\begin{equation}
\label{variance}
\chi(\tau,z_0) = \sum_{z_i}\left[\text{ \large T}\!\left\{
\tilde{Q}(z_i-z_0)\right\}-\tilde{Q}(z_i-\tau z_0)\right]^2,
\end{equation}
as a function of $\tau$ for different values of the ``energy
shift'' $z_0$. The sum Eq.~(\ref{variance}) was taken over
discrete set $z_i=-10+0.01i$ for $i=1,2, ..,220$. In
Fig.~\ref{diffz} we plot the variance for $z_0=1$ and $z_0=-1$.
Both curves have pronounced minima at $\tau=1.27$ and $1.22$,
respectively. This translates into the values of $\nu = 2.3 $ and
$\nu = 2.76$. Although these values are in good agreement with
known value of $\nu$, the accuracy of the above estimate is
limited.  The limitation is due to the fact that for $z_0=\pm 1$
the heights of maxima of the curves $\text{ \large
T}\!\left\{\tilde{Q}(z-z_0)\right\}$, shown in Fig. \ref{diffz},
differ from the height of $\tilde{Q}(z)$. This deviation would not
be a problem for smaller $z_0$. However for $z_0=\pm 0.5$ the
variance becomes small, and its dependence on $\tau$ becomes weak.
Apparently, the variance Eq.~(\ref{variance}) is affected by the
wiggles at the top of the curves $\text{ \large
T}\!\left\{\tilde{Q}(z-z_0)\right\}$ much stronger for $z_0=\pm
0.5$ than for $z_0=\pm 1$, which makes the evaluation of $\chi$
for small $z_0$ ineffective.

\section{conclusion}

It is interesting to point out that, while the classical limit of
the Chalker-Coddington model based on potential Eq.
(\ref{behaves}) reduces to the {\em bond} percolation, similar
form of potential Eq. (\ref{Random}) leads to the {\em site}
percolation. The reason is the symmetry of corresponding
potentials. As seen from Fig. \ref{trilat}, the hexagons
surrounding the nodes of potential Eq. (\ref{Random}) have common
{\em sides}. On the other hand, the squares, drawn around the
nodes of potential Eq. (\ref{behaves}) share the {\em vertices}.

Both the bond percolation on a square lattice and the site
percolation on triangular lattice have $p_c=1/2$, which is insured
by self-duality. The RG descriptions \cite{Klein} of both cases,
having fixed point, $p_c=1/2$, effectively preserve this
self-duality. As a result, the RG values for classical exponent
come out close to $\nu=4/3$ in both RG schemes. In this paper we
demonstrated that quantum extension of classical RG to the
triangular lattice also yields the critical exponent close to the
known value $\nu=2.33$.

The fact that the simple RG scheme, considered in the present
paper, describes the quantum Hall transition so accurately has a
deep underlying reason. Delocalized state in the quantum Hall
transition emerges as a result of competition of two trends: ({\em
i}) quantum interference processes that survive in the presence of
magnetic field (Hikami boxes, Ref. \onlinecite{hikami}) tend to
localize electron, while ({\em ii}) classical Lorentz force, by
causing electron drift, prevents it from repeating closed
diffusive trajectories. Both trends are incorporated into our RG
scheme. Obviously, chiral motion is the consequence of the Lorentz
force. Hikami boxes, on the other hand, are represented in the RG
step in the form of {\it figure-eight} trajectories, as
illustrated in Fig. \ref{contraction}.

Note finally, that simplicity of the RG description proposed here
suggest possibility to extend it to different from qunatum Hall
universality classes, see, {\em e.g.},
Refs.~\onlinecite{kagalovsky99,senthil99,gruzberg99,cardy}.

\section{Acknowledgements}
We are grateful to I. Gruzberg and V. Kagalovsky for numerous
discussions of the network models. This work was supported by the
BSF grant No. 2006201.

\appendix

\section{}

Here we elaborate on the mapping of the problem of percolation
over equipotential lines in the random potential Eq.
(\ref{Random}) and the conventional site percolation problem. The
easiest way to establish this mapping is to surround all sites of
triangular lattice with hexagons, as shown in Fig. \ref{trilat}.
If the site is occupied, then the hexagon is, say, black; if the
site is vacant, it is white. The distinctive property of the
triangular lattice is that, when two hexagons touch, they
automatically have a common side. Note in passing, that this is
not the case for a square lattice, where two squares, drawn around
the sites, may touch by sharing a vertex but not share a side.  In
the site-percolation problem, the bond between two neighboring
sites conducts if both of them are occupied. The same is to say
that conduction is possible between two touching hexagons, if they
are both black. Percolation threshold corresponds to the portion
of black hexagons when conduction over entire sample becomes
possible. The fact that  $p_c=1/2$ is a direct consequence of
geometrical arrangement of hexagons, due to which percolation over
black hexagons rules out the percolation over white hexagons.
Also, due to this arrangement, one of the colors always
percolates.

Consider now the potential Eq. (\ref{Random}). Black sites are now
those in which $V_0<\varepsilon$. Configuration of equipotentials
around this site is the rightmost of three shown in  Fig.
\ref{evolution}b. Accordingly, configuration of equipotentials
around the site with $V_0>\varepsilon$ is the leftmost of three
shown in  Fig. \ref{evolution}b. Consider now two neighboring
sites with $V_0<\varepsilon$. Fig. \ref{trilat} makes it apparent
that any two black points inside hexagons surrounding these sites,
are connected via black color. Thus, it terms of connectivity over
black, two neighboring sites with $V_0<\varepsilon$ are completely
similar to two neighboring black hexagons. Similarly, as can be
seen from Fig. \ref{trilat}, for two neighboring sites with
$V_0<\varepsilon$ and $V_0>\varepsilon$, the centers of
surrounding hexagons are disconnected. The same is true for two
neighboring hexagons of opposite colors in the percolation
problem. To complete the mapping, we note that, in percolation
problem, the connectivity of two neighboring hexagons depends {\em
entirely} on their colors, i.e., it does not depend on the color
of the other neighbors. In the same way, in the problem of
equipotentials, whether or not two neighboring sites are connected
is determined exclusively by the signs of $V_0-\varepsilon$ in
these sites.

\section{}

The form of the $S$-matrix Eq. (\ref{sm}) can be established with
the help of Fig. \ref{scatobj}. Matrix $S$ relates the incident,
$i_1$, $i_2$, $i_c$, and outgoing, $o_1$, $o_2$, $o_c$, amplitudes
via transmission coefficients, $x_1$, $x_2$, and $x_c$. The form
Eq. (\ref{sm}) follows from the system of six equations, which
include also the amplitudes $a_1$, $a_2$, and $a_c$ between the
points of close contact of corresponding equipotentials. As seen
from Fig. \ref{scatobj} the amplitudes $a_1$ and $a_2$ are related
via $x_1$ as
\begin{eqnarray}
\label{le1}
&&a_1=x_1a_2e^{i\phi_2}+y_1i_1,\nonumber\\
&&o_2=-y_1a_2e^{i\phi_2}+x_1i_1.
\end{eqnarray}
The amplitudes $a_c$ and $a_1$ are related via $x_c$
\begin{eqnarray}
\label{le2}
&&a_c=x_c\,a_1e^{i\phi_1}+y_c\,i_c,\nonumber\\
&&o_1=-y_c\,a_1e^{i\phi_1}+x_c\,i_c.
\end{eqnarray}
Finally, the amplitudes $a_2$ and $a_c$ are related via $x_2$
\begin{eqnarray}
\label{le3}
&&a_2=x_2a_c\,e^{i\phi_c}+y_2i_2,\nonumber\\
&&o_c=-y_2a_c\,e^{i\phi_c}+x_2i_2.
\end{eqnarray}
In Eqs. (\ref{le1})-(\ref{le3}), the phases $\phi_1$, $\phi_2$,
and $\phi_c$ are the Aharonov-Bohm phases accumulated,
respectively, by waves $a_1$, $a_2$, $a_c$ between the points of
closed contact. Excluding $a_1$, $a_2$, $a_c$ from Eqs.
(\ref{le1})-(\ref{le3}) we recover $S$-matrix Eq. (\ref{sm}), in
which $x_3$ stands for $x_c$ and $\phi_3$ for $\phi_c$.

In order to derive Eq. (\ref{i2o2}) we set in Eqs.
(\ref{le1})-(\ref{le3}) $i_2=o_2e^{i\varphi_2}$, as enforced by a
contraction in Fig. \ref{contraction}a. Upon setting $i_1=0$, we
find the proportionality coeffcient between $o_1$ and $i_c$. This
recovers Eq. (\ref{i2o2}), in which $\varphi=\varphi_2+\phi_2$.
Similarly, Eq. (\ref{o2i1}) is recovered upon setting
$i_1=o_2e^{i\varphi}$, as shown in Fig. \ref{contraction}b, and
relating the amplitudes $o_1$ and $i_c$.

\section{}

%%%%%%%%%%%%%%%%%%%
\begin{figure}[t]
\centerline{\includegraphics[width=90mm,angle=0,clip]{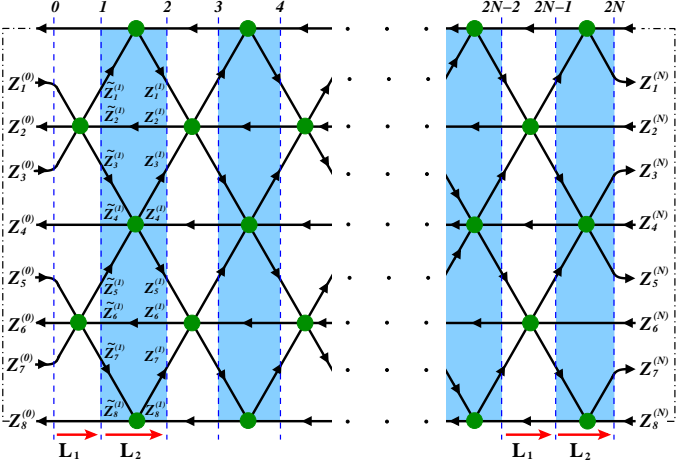}}
\caption{(Color online) A slice of a network Fig. \ref{network} of
width $M=8$ is shown. Three amplitudes on the links to left of
green dot are related to three amplitudes to the right of green
dot via  matrix $\mathbf{X}$. The upper and the lower boundaries
of the slice are connected by dashed-dotted lines manifesting that
the amplitudes on these boundaries are the same by virtue of
cyclic boundary conditions. Upon passing the white stripe, the
vector of the amplitudes $\{Z_i\}$ is multiplied by the matrix
$\mathbf{L}_1$.
 Upon passing the blue stripe the vector $\{\tilde{Z}_i\}$
is multiplied by $\mathbf{L}_2$.} \label{newNM}
\end{figure}
%%%%%%%%%%%%%%%%%%
In this Appendix we illustrate the transfer-matrix method in
application to the network model on triangular lattice. Consider,
by analogy to Ref. \onlinecite{CC}, a slice of length, $N$, shown
in Fig. \ref{newNM}. It contains $M=8$ incident links, $Z_i^{(0)}$,
and  $8$ outgoing links, $Z_i^{(N)}$. In general, the number of
links must be $M= 4m$, where $m$ is integer. The slice shown
in Fig.  \ref{newNM} can be obtained from the general network
Fig. \ref{network} by two vertical cuts through the centers of
triangles. The bottom links $Z_8^{(0)}$ and $Z_8^{(N)}$ in
Fig. \ref{newNM} are connected to the corresponding top links
by dashed lines, reflecting the fact that these links must be
identified with each other in order to impose a periodic boundary
condition \cite{CC}.
Scattering of waves at each node is described by the matrix
$S=S_\vartriangle$, Eq. (\ref{sc}), which relates the amplitudes
$(o_1,o_2,o_3)$ to $(i_1,i_2,i_3)$. To adapt $S=S_\vartriangle$  to
the  transfer-matrix algorithm, one has to recast Eq.~\ref{ss1}
into the form
\begin{eqnarray}
\label{ss2}
&&\left(\begin{array}{c} o_2\\
i_2\\ o_3
\end{array}\right) = \mathbf{X}\left(\begin{array}{c} i_1\\
o_1\\ i_3
\end{array}\right),
\end{eqnarray}
which connects the amplitudes to the left and to the right
from the node. Direct calculation yields the following
form of matrix $\mathbf{X}$ in terms of elements, $s_{ij}$
of $S_\vartriangle$.
\begin{eqnarray}
\label{ns1} \mathbf{X}=\left(\begin{array}{ccc}
\left[s_{21}-\frac{s_{11}s_{22}}{s_{12}}\right]&\frac{s_{22}}{s_{12}}
&\left[s_{23}-\frac{s_{22}s_{13}}{s_{12}}\right]\\
\,\\
-\frac{s_{11}}{s_{12}}&\frac{1}{s_{12}}&-\frac{s_{13}}{s_{12}}\\
\,\\
\left[s_{31}-\frac{s_{11}s_{32}}{s_{12}}\right]&\frac{s_{32}}{s_{12}}
&\left[s_{33}-\frac{s_{32}s_{13}}{s_{12}}\right]
\end{array}\right).
\end{eqnarray}
For energies close to the critical $\varepsilon=0$, the $S$-matrix
is given by Eq.~(\ref{sc}); $S=S_\vartriangle$. Making use of
Eq.~(\ref{sc}), it is straightforward to find the energy dependence
of $\mathbf{X}$
\begin{eqnarray}
\label{ns2} \mathbf{X}= \left(\begin{array}{ccc}
2(1+\varepsilon)&-2(1+\varepsilon)&1\\
%\,2/3&-1/3&\,2/3\\
\,\\
2(1+\varepsilon)&-3&2(1-\varepsilon)\\
%\,2/3&\,2/3&-1/3\\
\,\\
1&-2(1-\varepsilon)&2(1-\varepsilon)
%-1/3&\,2/3&\,2/3
\end{array}\right).
%\nonumber
\end{eqnarray}

Operators $\mathbf{L}_1$ and $\mathbf{L}_2$ act in "white" and
"blue" stripes, respectively. Operator $\mathbf{L}_1$ performs the
transformation of the vector of amplitudes, $\{Z_i^{(n)}\}$, into
$\{\tilde{Z}_i^{(n+1)}\}$, while $\mathbf{L}_2$ performs the
transformation of $\{\tilde{Z}_i^{(n)}\}$ into $\{Z_i^{(n)}\}$.
Matrix forms of $\mathbf{L}_1$ and $\mathbf{L}_2$ in terms of
elements of matrix, $\mathbf{X}$, are the following
\begin{eqnarray}
\mathbf{L}_1 =\left( \begin{array}{cccc|cccc}
x_{11}& x_{12}& x_{13}& 0 &\cdot &\cdot &\cdot &\cdot \\
x_{21}& x_{22}& x_{23}& 0 &\cdot &\cdot &\cdot &\cdot \\
x_{31}& x_{32}& x_{33}& 0 &\cdot &\cdot &\cdot &\cdot \\
0&0 &0 &1 &\cdot &\cdot &\cdot &\cdot  \\
\hline
\cdot &\cdot &\cdot &\cdot &x_{11}& x_{12}& x_{13}& 0 \\
\cdot &\cdot &\cdot &\cdot &x_{21}& x_{22}& x_{23}& 0 \\
\cdot &\cdot &\cdot &\cdot &x_{31}& x_{32}& x_{33}& 0 \\
\cdot &\cdot &\cdot &\cdot &0&0 &0 &1
\end{array} \right),
%\nonumber
\end{eqnarray}
\begin{eqnarray}
\mathbf{L}_2 =\left( \begin{array}{cc|cccc|cc}
 x_{33}&0 &\cdot &\cdot &\cdot &\cdot & x_{31}& x_{32} \\
0&1&\cdot &\cdot &\cdot &\cdot &0&0 \\
\hline
\cdot &\cdot & x_{11}& x_{12}& x_{13}& 0 &\cdot &\cdot \\
\cdot &\cdot & x_{21}& x_{22}& x_{23}& 0 &\cdot &\cdot \\
\cdot &\cdot & x_{31}& x_{32}& x_{33}& 0 &\cdot &\cdot \\
\cdot &\cdot &0 &0 &0 &1 &\cdot &\cdot   \\
\hline
x_{13}&0 &\cdot &\cdot &\cdot &\cdot & x_{11}& x_{12} \\
x_{23}&0 &\cdot &\cdot &\cdot &\cdot& x_{21}& x_{22}
\end{array} \right),
%\nonumber
\end{eqnarray}
with dots standing for zeroes. Specific form of $\mathbf{L}_2$
accounts for the cyclic boundary conditions in the vertical
direction.

In addition to the scattering at the nodes (green dots
in Fig.~\ref{newNM}), propagation along the network links
accumulates  random phases. If we denote the
phases on the links crossing vertical lines $2n$ by
$\varphi_i^{(n)}$ and those crossing lines $2n+1$ by
$\psi_i^{(n)}$, then all the random phases at the links can be
taken into account by introducing diagonal phase matrices
\begin{eqnarray}
\mathbf{P}_1^{(n)}=\text{diag}\{e^{i\varphi_1^{(n)}},
\cdots,e^{i\varphi_8^{(n)}}\}
\end{eqnarray}
and
\begin{eqnarray}
\mathbf{P}_2^{(n)}=\text{diag}\{e^{i\psi_1^{(n)}},
\cdots,e^{i\psi_8^{(n)}}\}.
\end{eqnarray}
Finally, the transfer matrix,
$\mathbf{T}$, of the slice Fig.~\ref{newNM} is given by the product
\begin{eqnarray}
\label{tm}
\mathbf{T}=\prod_{n=N-1}^{0}\mathbf{L}_2\mathbf{P}_2^{(n)}\mathbf{L}_1
\mathbf{P}_1^{(n)},
\end{eqnarray}
which runs in the reverse order. This product is completely
analogous to the transfer matrix of the slice in the
Chalker-Coddington model.

\end{document}